\documentclass[12pt,letterpaper]{article}

\usepackage{graphicx}
\usepackage{amssymb}
\usepackage{geometry}
\usepackage{color}
\usepackage{amsmath}
\usepackage{amssymb}
\usepackage{epstopdf}
\DeclareGraphicsExtensions{.eps}

\usepackage{setspace}

\usepackage[T1]{fontenc}
\usepackage[latin9]{inputenc}

\usepackage{newunicodechar}
\usepackage{lineno}
\usepackage{fancyhdr}
\usepackage{authblk}

\usepackage{natbib}

  \makeatletter
\renewcommand*{\@fnsymbol}[1]{\ensuremath{\ifcase#1\or \dagger\or*\or  \ddagger\or \mathsection\or \mathparagraph\or \|\or **\or \dagger\dagger \or \ddagger\ddagger \else\@ctrerr\fi}}
\makeatother
    
\title{Efficient MCMC implementation of multi-state mark-recapture models}
\author[1]{Ford, J.H\thanks{Corresponding author: Jessica.Ford@csiro.au}}
\author[1]{Patterson, T.A.}
\author[1]{and Bravington, M.V.}
\affil[1]{CSIRO, Castray Esplanade, Hobart, 7001, TAS, Australia.} 

\date{\today}

\begin{document}

\begin{titlepage}

\maketitle

\begin{abstract}
Inherent differences in behaviour of individual animal movement can
introduce bias into estimates of population parameters derived from
mark-recapture data. Additionally, quantifying individual heterogeneity
is of considerable interest in it's own right as numerous studies
have shown how heterogeneity can drive population dynamics. In this
paper we incorporate multiple measures of individual heterogeneity
into a multi-state mark-recapture model, using a Beta-Binomial Gibbs
sampler using MCMC estimation. We also present a novel Independent
Metropolis-Hastings sampler which allows for efficient updating of
the hyper-parameters which cannot be updated using Gibbs sampling.
We tested the model using simulation studies and applied the model
to mark-resight data of North Atlantic humpback whales observed in
the Stellwagen Bank National Marine Sanctuary where heterogeneity
is present in both sighting probability and site preference. Simulation
studies show asymptotic convergence of the posterior distribution
for each of the hyper-parameters to true parameter values. In application
to humpback whales individual heterogeneity is evident in sighting
probability and propensity to use the marine sanctuary. \medskip{}


\end{abstract}

\end{titlepage}

\section{Introduction}

Mark-recapture analysis is a fundamental tool for understanding populations.
Demographic parameters estimated from mark-recapture data, such as
survival and reproduction, are used to infer population status and
predict future dynamics of population. Standard statistical techniques
for estimating population parameters from mark-recapture data, often
by omission, implicitly assume animals to be identical. This is something
shared with most ecological models which assume populations to be
composed of identical average individuals \citep{Grimm2002}. While
this is a convenient and often necessary simplification, individuals
in wild populations typically do not behave, grow or reproduce identically.
Neither are they identically observed by researchers \citep{Crespin2008}.
This heterogeneity between individuals presents a challenge both for
the collection and analysis of mark-recapture data. Individuals in
natural populations tend to exhibit substantial individual variation
which can manifest through demographic parameters \citep{Lebreton1992}.
For example, inherent individual differences in movement and behavior
can introduce bias into mark-recapture-derived estimates; most notoriously
for inferring population size \citep{Pledger2010}. Moreover characterizing
the extent of individual variability in some aspect of individuals
biology is often of considerable interest in it's own right as it
can have substantial impact on population \citep{Franklin2000}
and even ecosystem dynamics \citep{Vieilledent2010}.\medskip{}

While mark-recapture analysis methods capable of explicitly including
individual heterogeneity are not the norm, several studies have addressed
the problem. \citet{Pledger2000} and \citet{Norris1996} developed
finite mixture models which assume that differences between individuals
can be explained by categorising into a finite set of latent groups.
Despite the computational advantage over other methods such as continuous
random effects \citep{Ford2012}, this approach relies on the assumption
of a prespecified number of groups and can result in model selection
issues such as determining the number of groups of individuals sharing
the same survival or detection parameter \citep{Cubaynes2012}. \medskip{}

Individual-level random effects using continuous distributions are
a natural candidate for modelling heterogeneity as they do not require
that individuals be categorized into prespecified groups. Random effects
are useful as they can allow a proportion of variance on some population
parameter (i.e. detection) to be related to persistent unobserved
individual heterogeneity. In this context random effects allow individuals
to obtain their own instance of a population parameter (e.g. detection
probability), drawn from a common distribution whose parameters are
estimated. However, despite inclusion of random effects being the
focus of much research \citep{Maunder2008,Lebreton2009,Barry2003,Royle2008,Huggins2001,Burnham1978,Gimenez2010},
methods for the inclusion of multiple individual-level continuous
random effects\emph{ }into mark-recapture models are still lacking.
This is in part due to the complex calculation required to solve for
these random effects. \medskip{}

To estimate random effects models it is necessary to integrate across
all possible values of the individual-level random effects which is
not straightforward. \citet{Ford2012} provide one solution using
the open source software Automatic Differentiation Model Builder with
Random Effects (ADMB-RE, \citealt{Fournier2012}). Another possible
technique is the use of Markov chain Monte Carlo (MCMC) as this provides
a solution to the calculation of marginal distributions which involve
complex integrals \citep{Gilks1996}. 

\medskip{}
One method to capture heterogeneity is through the development of
multi-state mark-recapture models. Multi-state mark-recapture models,
first developed by Arnason \citep{Arnason1972,Arnason1973}, extend
traditional mark-recapture models by allowing animals to be in different
\textquoteleft{}states\textquoteright{} \citep{Lebreton2009}. The
'movement' parameter, the probability of transitioning between states,
was originally introduced to distinguish between emigration and mortality.
The states, which may or may not be directly observable, can include,
but are not limited, to breeding, location, and behaviour. State can
affect the probability of observation, and this can be built into
the multi-state framework \citep{Lebreton2002}. These models can
be extended \citep{Pradel2005} through the use of a hidden Markov
model framework, which incorporate a more realistic assessment of
natural events as transitions between states can be treated as a Markov
process that is not directly observable \citep{Conn2009,Zucchini2008}.
In hidden Markov models, two time series - the observation and process
components - run in parallel \citep{Gimenez2012}. The observation
process (e.g. seen/not seen) does not usually reveal the current underlying
state directly, but does provide indirect information on the probable
state (e.g. present in the marine sanctuary, not present, or dead).
Modelling both the process and observation component enables the separation
of the real signal from the observation error \citep{Patterson2008}.
\medskip{}

Here we develop a multi-state model using a hidden Markov model framework
which explicitly includes individual heterogeneity in sighting probability
and site fidelity, and we apply it to a long term data set of North
Atlantic humpback whales at the Stellwagen Bank National Marine Sanctuary
(SBNMS) off the northeast coast of the United States. Individual humpback
whales have been intensively studied in this region since the late
1970s. However, the SBNMS covers only a small part of the population's
summer range, and although some individuals are seen regularly there,
none are thought to remain permanently within its boundaries. This
presents a challenge when studying the vital rates of whales using
the area and the effectiveness of management initiatives.\medskip{}

There are presently over 500 marine protected areas (MPAs) primarily
for marine mammals. Declines in marine mammal populations have been
attributed to a variety of factors: entanglement in fishing gear \citep{Hoyt2011,Johnson2005,DAgrosa2000},
over fishing \citep{Read2006}, pollution \citep{Wilgart2007}, and
ship strikes \citep{Laist2001,Knowlton2001}. As these threats are
often concentrated in space, MPAs have been advocated as an effective
management strategy for mitigation. Determining the effectiveness
of a MPA's spatial configuration (e.g. location, extent) is challenging.
An often overlooked factor is how individual heterogeneity in spatial
use can mediate the effectiveness of a MPA. \medskip{}

Heterogeneity in sighting probability is a well known phenomenon \citep{Hammond1986,Hammond1990}
as the probability of sighting relies on individual behaviour at the
beginning of a dive. The angle to which an individual's fluke shows
on diving determines the probability of a successful photograph. The
probability of being seen and recognized is therefore a combination
of the true observation error (e.g. the randomness in viewing flukes)
and also the real biological signal arising from individual heterogeneity
in presence and absence in the SBNMS. The difficulty lies in determining
the underlying behaviour of the whales: observations are indications
of presence in the SBNMS but whales may also be present and not observed.

\medskip{}

In this paper we develop a Bayesian hierarchical approach to incorporating
random effects into a hidden Markov model using MCMC estimation. Following
work by \citet{Zucchini2008} and \citet{Scott2002} we develop a
Beta-Binomial Gibbs sampler for the hidden Markov model. We also present
a novel Independent Metropolis-Hastings sampler which allows efficient
updating of the hyper-parameters which cannot be updated using Gibbs
sampling. We employ a hidden Markov model which allows for individual
variability on the probability of observation and the probability
of remaining in either of the two states: resident in the SBNMS or
elsewhere \cite[see][]{Ford2012}. A two-state model is used for
simulation testing and we test for asymptotic convergence of individual
parameter values and population-level hyper-parameters using the two-state
model. The full three-state model including death \cite[see][]{Ford2012},
is applied to a data from North Atlantic Humpback whales. \medskip{}

\section{Methods \label{sec:Model}}

\subsection{Two-state model used for simulations}

This section describes our model for simulated data. Although it is
closely aligned to the real SBNMS situation, analyzing the real data
entails attention to a few extra details, omitted here for clarity.\medskip{}

Hidden Markov models or multi-state models are split into a process
and an observation model. For the process model, we assume that at
time $t$ an animal $i$ can be in either of two states $S_{it}$:
Here and Away, or H/A for short. Changes in the state over time are
governed by a Markov process with transition matrix $\gamma$, so
(omitting dependence on $i$ for now) for any two states $s$ and
$s^{*}$ we have
\begin{eqnarray*}
\mathbb{P}\left[S_{t+1}=s^{*}\right] & =\sum_{s} & \gamma^{ss*}\mathbb{P}\left[S_{t}=s\right]
\end{eqnarray*}
\medskip{}

The four elements of $\gamma$ can be written in terms of just two
parameters $\gamma^{HH}$ and $\gamma^{AA}$ (respectively the probabilities
of staying Here and staying Away), as follows:

\begin{eqnarray*}
\gamma & = & \begin{pmatrix}\gamma^{HH} & \left(1-\gamma^{HH}\right)\\
\left(1-\gamma^{AA}\right)\, & \gamma^{AA}
\end{pmatrix}
\end{eqnarray*}

\medskip{}

For the observation model, there are \textquotedblleft{}capture attempts\textquotedblright{}
(photo-ID expeditions) at each $t$, in which an animal may be seen
if and only if it is Here. Our data for animal $i$ are thus a time
series $X{}_{i,t_{1i}:T}$ of $0$ s (not seen) and $1$ s (seen)
where $t_{1i}$ denotes the first observation of the animal (see below)
and $T$ the most recent expedition. If $X_{it}=1$ then we know $S_{it}=\mbox{H}$
, but if $X_{it}=0$ the state cannot be determined for certain. Formally,
the probability of observation given state is expressed in terms of
a parameter $\pi$ by 

\begin{eqnarray*}
\mathbb{P}\left[X_{it}=1|S_{it}=s\right] & = & \left\{ \begin{array}{cc}
\pi_{it} & s=\mbox{H}\\
0 & s=\mbox{A}
\end{array}\right.\\
\mathbb{P}\left[X_{it}=0|s\right] & = & 1-\mathbb{P}\left[X_{it}=1|s\right]
\end{eqnarray*}
\medskip{}

We start each animal's series at its first sighting of the given year,
and condition on $S_{t_{1i}i}=1$ . In the synthetic data used in
this paper, we assume no recruitment and simulate data with all animals
present and seen on the first occasion. 

\medskip{}

\subsection{Computation\label{sec:Computation}}

Given a two-state hidden Markov model with underlying latent state
chain, the probability of observation is denoted by $\pi$, and the
transition probability matrix of the hidden Markov chain by $\gamma$.
Given a series of observations $X_{1:T}$ and prior distributions
on $\pi$ and $\gamma$, our aim is to estimate the posterior distribution
using MCMC. The MCMC routine developed in this paper involves four
main steps (five in application to real data).

\medskip{}

The MCMC algorithm for one iteration consists of the following steps:
\begin{enumerate}
\item Sampling the hidden state chain for all individuals.
\item Calculating summary statistics per individual conditional on its sampled
states.
\item Updating the posteriors for individual-level parameters $\pi_{i}$,
$\gamma_{i}^{HH}$ and $\gamma_{i}^{AA}$ separately using Gibbs sampling
from Beta distributions.
\item Updating the population-level hyper-parameters $\pi$, $\gamma^{HH}$
and $\gamma^{AA}$ using an Independent Metropolis-Hastings sampler
with three proposal distributions whose parameters vary across iterations.
\item Updating population-level fixed effects using an Independent Metropolis-Hastings
sampler with a fixed proposal distribution: a multivariate t-distribution
whose mean and variance are set using a preliminary fit from ADMB
\cite[see][]{Ford2012}. 
\end{enumerate}

\subsection{Forward-Backward recursion\label{sub:Forward-Backward-recursion}}

\textcolor{black}{In order to update individual-level parameter values
($\theta_{i}$ for individual-level values and $\theta$ for population-level
values) at each iteration, we require counts of successes and trials
for each individual. Following Zucchini and MacDonald \citeyearpar{Zucchini2009}
we simulate a sample path ($Z^{(T)}$) from the conditional distribution}

\[
\mathbb{P}(Z^{(T)}|x^{(T)},\theta)=\mathbb{P}(Z_{T}|x^{(T)},\theta)\times\prod_{t=1}^{T-1}\mathbb{P}(Z_{t}|x^{(T)},Z_{t+1}^{T},\theta)
\]

\textcolor{black}{We draw the sample path in the order $Z_{T},Z_{T-1},...,Z_{1}$.
To do this we need }

\[
\mathbb{P}(Z_{t}|x^{(t)},\theta)=\frac{\mathbb{P}(Z_{t},x^{(t)}|\theta)}{\mathbb{P}(x^{(t)}|\theta)}=\frac{\alpha_{t}(Z_{t})}{L_{t}}\propto\alpha_{t}(Z_{t})\text{ , for }t=1,...,T.
\]

\textcolor{black}{where $\alpha_{t}$ are the forward probabilities.
Given the forward probabilities we backward sample ($T,T-1,...,1$)
the hidden state chain. This recursion scheme is referred to as the
Forward-Backward recursion scheme \citep{Scott2002,Zucchini2009}.
The counts are then obtained from these sampled state chains. This
recursion scheme consists of one forward pass and one backward pass,
per individual, for each iteration of the MCMC sampler. The Forward
recursion produces the forward probability vector }$\alpha_{2},...,\alpha_{n}$,
containing the probabilities of the underlying hidden states for each
observation given all observed data up to time $t$. We calculate
these forward probabilities, from $1:T$, for each state, given the
observed data ($X$). 

\begin{eqnarray*}
\alpha_{t}(S_{t}) & = & \mathbb{P}(S_{t}|X_{1:t})\\
 & = & \sum_{S_{t-1}}\mathbb{P}(S_{t-1}|X_{1:t-1})\mathbb{P}(S_{t}|S_{t-1})\mathbb{P}(X_{t}|S_{t})\\
 & = & \sum_{S_{t-1}}\alpha_{t-1}(S{}_{t-1})\mathbb{P}(S_{t}|S_{t-1})\mathbb{P}(X_{t}|S_{t})
\end{eqnarray*}

where $\mathbb{P}(X_{t}|S_{t})$ denotes the probability of the data
given the state. The backward pass generates a sample path $Z^{(T)}$
of the hidden state chain in the order $t=T,T-1,T-2,...,1$, making
use of the following proportionality argument:

\begin{equation}
\mathbb{P}(Z_{t}|x^{(T)},Z_{t+1}^{T},\theta)\propto\alpha_{t}(Z_{t})\mathbb{P}(Z_{t+1}|Z_{t},\theta).\label{eq:sth column}
\end{equation}
 The second factor in equation \ref{eq:sth column} is simply a one-step
transition probability in the Markov chain.

\subsection{Updating individual values of $\gamma^{HH}$, $\gamma^{AA}$ and
$\pi$\label{sub:Posterior-sampling-by}}

Observations for an individual are assumed Binomial with probability
$\pi_{i}$. As the Beta prior for $\pi$ is conjugate to the Binomial,
the posterior is also Beta. For the probability of observation there
is a trial whenever an animal is Here; the outcome is whether it was
or wasn't seen. There is no trial when then animal is Away, since
it is then guaranteed not to be seen. The summary statistics for the
transition probabilities ($\gamma^{HH}$ and $\gamma^{AA}$) are calculated
from the sampled state chains. For $\gamma^{HH},$ there is a trial
whenever the animal was Here (excluding the final period); the outcome
is whether it stayed Here or not. A similar scheme applies to $\gamma^{AA}$.\medskip{}

Following the hidden Markov model Forward-Backward recursion scheme
and calculation of the successes and trials, we update individual-level
random effects by sampling from the posterior distribution. Each individual-level
parameter $\theta_{i}$ ($\pi_{i}$, $\gamma_{i}^{HH}$ or $\gamma_{i}^{AA}$
) is an independent sample from a prior Beta distribution: $\theta_{i}\sim\mbox{Beta}(a,b)$. 

\medskip{}

The joint posterior is

\begin{eqnarray*}
p(\theta,a,b|y) & \propto & p(a,b)p(\theta|a,b)p(y|\theta)\\
 & \propto & p(a,b)\prod_{i=1}^{N}\frac{\Gamma(a+b)}{\Gamma(a)\Gamma(b)}\theta_{i}^{a-1}(1-\theta_{i})^{b-1}\prod_{i=1}^{N}\theta_{i}^{y_{i}}(1-\theta_{i})^{n_{i}-y_{i}}
\end{eqnarray*}

Gibbs sampling can be used to update $\theta_{i}$, since the full
conditional for $\theta$ is available: $\theta_{i}|a,b,y\sim Beta(y_{i}+a,\, n_{i}-y_{i}+b)$
where $y_{i}$ indicates the number of successes (e.g. number of observations)
for individual $i$, and $n_{i}$ the number of trials. 

\medskip{}

\subsection{An Independent Metropolis-Hastings sampler - the choice of proposal
distribution for $a$ and $b$ \label{sub:An-independent-metropolis-hastin}}

Extending the hierarchy above to deal with the population-level hyper-parameters
$a$ and $b$ we have 

\begin{eqnarray*}
y_{i}|\theta_{i} & \sim & Bin(\theta_{i},n_{i})\\
\theta_{i}|a,b & \sim & Beta(a,b)\\
a,b & \sim & p(a,b)
\end{eqnarray*}
where $p(a,b)$ indicates the prior distribution for the hyper-parameters.
Given the joint posterior distribution of parameters is $p(\theta,a,b|y)\propto p(a,b)p(\theta|a,b)p(y|\theta)$
we can see that given $\theta$, the dependency on the data disappears
for the hyper-parameters. Thus in order to update our population-level
hyper-parameters, we require only the updated values of $\theta$.
As there is no conjugate prior, these population-level hyper-parameters
are updated using an Independent  Metropolis-Hastings sampler. \medskip{}

The Independent Metropolis-Hastings sampler works by ignoring the
current value $\theta^{*}$, and sampling the candidate value for
update, $\theta\newunicodechar{°}{\degree}$, directly from a proposal distribution
$\tilde{f}$ that should be close to the ideal distribution $f\left(\theta\right)$.
Following this, the acceptance ratio becomes
\[
\frac{f\left(\theta^{*}\right)}{f\left(\theta\newunicodechar{°}{\degree}\right)}\frac{\tilde{f}\left(\theta\newunicodechar{°}{\degree}\right)}{\tilde{f}\left(\theta^{*}\right)}
\]
 which does not completely cancel. However, insofar as the approximating
distribution is close to the target distribution, the average acceptance
ratio will be close to $1$. \medskip{}

The logit scale is used, since the distribution of $\mbox{logit\ensuremath{\theta}}$
is reasonably Normal for any reasonable $Beta$ prior on $\theta$
(i.e. unless $\left(a,b\right)$ have become extreme). The collection
of $\mbox{logit\ensuremath{\theta}}$ is distributed approximately
Normal, $N(\mu,\sigma^{2})$. We need to update this distribution
with the collection of individual $\theta_{i}$ to get a posterior
for $\mbox{logit\ensuremath{\theta}}$ - $N(\mu^{\prime},\sigma^{2\prime})$.
In order to do this we need to turn $(\mu^{\prime},\sigma^{2\prime})$
into the corresponding $(a,b)$. There is a simple relationship between
$\left(a,b\right)$ and $\left(\mu,\sigma^{2}\right)$, the mean and
variance of $\mbox{logit\ensuremath{\theta}}$. Using the cumulant
generating function $K(t)$ we have

\begin{eqnarray*}	
\mathbb{E}\left[\exp\left(t\mbox{logit}p\right)\right] & = & \frac{1}{B\left(a,b\right)}\int\exp\left(t\log p-t\log\left(1-p\right)\right)p^{a-1}\left(1-p\right)^{b-1}dp\\
 & = & \frac{1}{B\left(a,b\right)}\int p^{a+t-1}\left(1-p\right)^{b-t+1}dp\\
 & = & \frac{B\left(a+t,b-t\right)}{B\left(a,b\right)}\\
 & \implies & K\left(t\right)=\log\mathbb{E}\left[\exp\left(t\mbox{ logit}p\right)\right]\\
 &  = &  \log\Gamma\left(a+t\right)+\log\Gamma\left(b-t\right)-\log\Gamma\left(a+b\right)-\log\Gamma\left(a\right)- \\
 & & \log\Gamma\left(b\right)+\log\Gamma\left(a+b\right)\\
 & \implies & K'\left(t\right)=\psi\left(a+t\right)-\psi\left(b-t\right)\\
 &  & K''\left(t\right)=\psi'\left(a+t\right)+\psi'\left(b-t\right)\\
 & \implies & \mu=\mathbb{E}\left[\mbox{logit}p\right]=K'\left(0\right)=\psi\left(a\right)-\psi\left(b\right)\\
 &  & \sigma^{2}=\mathbb{V}\left[\mbox{logit}p\right]=K''\left(0\right)=\psi'\left(a\right)+\psi'\left(b\right)\\
\\
 &  & \mbox{where }\psi(x)=\frac{d}{dx}log\Gamma(x)
\end{eqnarray*}

where $\psi$ indicates a moment; the first moment, $\psi$, is the
mean and the second moment, $\psi'$, the variance. To get $(a,b)$
from ($\mu,\sigma^{2}$) we apply a Newton-Raphson iteration to 

\begin{eqnarray*}
\mu & = & \psi(a)-\psi(b)\\
\sigma^{2} & = & \psi'(a)+\psi'(b).
\end{eqnarray*}
Given approximate starting values, $\mu=ln(a)/ln(b)$ and $\sigma^{2}=ln(a+b)$,
we apply a Newton-Raphson iteration to solve for $\left(a,b\right)$.
For a given $\left(\mu,\sigma^{2}\right)$, we seek
\begin{eqnarray*}
\left(a,b\right) & \mbox{{s.t.}} & F\left(\theta\right)=\left[\begin{array}{c}
\psi\left(a\right)-\psi\left(b\right)\\
\psi'\left(a\right)+\psi'\left(b\right)
\end{array}\right]-\begin{bmatrix}\mu\\
\sigma^{2}
\end{bmatrix}=0.
\end{eqnarray*}
 Thus
\begin{eqnarray*}
F'\left(\theta\right) & = & \begin{bmatrix}\psi'\left(a\right) & -\psi'\left(b\right)\\
\psi''\left(a\right) & \psi''\left(b\right)
\end{bmatrix}
\end{eqnarray*}
and we update via $\theta^{r+1}=\theta^{r}-\left[F'\left(\theta^{r}\right)\right]^{-1}F\left(\theta^{r}\right)$,
where $r+1$ indicates the next iteration. \medskip{}

For the approximating distribution ($\tilde{f}$), we assume a vague
conjugate-prior for the mean and variance $\left(\mu,\sigma^{2}\right)$
with the following conjugate hyper-priors for the $\mbox{logit}\theta_{i}$

\begin{eqnarray*}
\mbox{logit}\theta_{i}|\mu,\tau & \sim & N(\mu,\tau)\\
\mu|\tau & \sim & N(\mu_{0},n_{0},\tau)\\
\tau & \sim & Ga(\alpha_{\tau},\beta_{\tau})
\end{eqnarray*}
where $\mu_{0}=0$, $n_{0}\tau=0.1$, $\alpha_{\tau}=0.1$ and $\beta_{\tau}=0.1$.
Given the collection of $\mbox{logit}\theta_{i}$ we update this to
a conjugate posterior $\tilde{f}\left(\mu,\sigma^{2}|\mbox{logit}\theta_{i}\right)$
in the standard way for conjugate Gaussian problems. Following this
we sample $\theta^{*}=\left(\mu^{*},\sigma^{2*}\right)$ from $\tilde{f}\left(.\right)$
and compute $\tilde{f}\left(\theta^{*}\right)$; then $\left(\mu^{*},\sigma^{2*}\right)$
are back-transformed to $\left(a^{*},b^{*}\right)$. Following this
we compute the (vague prior times) log-likelihood of the collection
of $\theta_{i}$ under $\left(a^{*},b^{*}\right)$ and current values;
and finally the acceptance ratio is calculated and the hyper-parameters
are updated accordingly. 

\medskip{}

\subsection{Updates to fixed effects \label{sub:Updates-to-fixed-1}}

Although not used in simulations, the inclusion of any fixed effects
in the model are updated using an unchanging Independent Metropolis-Hastings
sampler. For example: for individual $i$, given a population-level
fixed effect $b$, individual-level effect $\alpha_{i}$, and a design
matrix $X$, the individual-level parameter of interest $\psi_{i}$,
is calculated as $\mbox{logit}\psi_{i}=\alpha_{i}+X_{i}b$. If $\alpha_{i}$
are sampled from a Beta distribution then these individual-level effects
are first transformed using the logit-link. \medskip{}

An Independent Metropolis-Hastings sampler is used to update these
values as there is no conjugate prior and, unlike in update to the
population hyper-parameters, it is not obvious how to generate an
approximate adaptive proposal. The use of an Independent Metropolis-Hastings
sampler avoids the need to adjust the tuning parameters required in
a random walk MCMC and the Laplace approximation, results from ADMB,
provides a good approximation to the posterior for these parameters
which can be used to design the unchanging proposal distribution.\medskip{}

The update to population-level parameters can be made using a multivariate
t-distribution with 5 degrees of freedom ($\mbox{MVT}{}_{5}$) with
covariance structure based on preliminary results from ADMB. The same
structure and data was used for the model in ADMB and the results
were used to facilitate block updates from the multivariate t-distribution.
In comparison to a multivariate Normal proposal, the multivariate
t-distribution allows for the potential of thick tails in the posterior
and better mixing of the chain. The use of a multivariate distribution
is more efficient than block updates of univariate distributions when
there is strong correlation between parameters and results in an efficient
proposal distribution which is close to the expected posterior estimates.
The log-likelihood is used for computations in the Metropolis-Hastings
acceptance ratio to avoid potential numerical instabilities which
may occur with use of the likelihood.

\subsection{Application to North Atlantic Humpback whales\label{sec:Application-to-North}}

A three-state hidden Markov model including death, developed in \citet{Ford2012},
was applied to a sub-set of 176 North Atlantic Humpback whales sighted
more than 30 times in the SBNMS between 1979 and 2005. 

Individual-level random effects were included on each of $\pi$, $\gamma^{HH}$
and $\gamma^{AA}$ and are updated exactly as described for the two-state
model applied to synthetic data. We assume individual-level parameters
to be consistent over time but have allowed for population-level annual
variation in probability of remaining Here using logit-links: $\mbox{logit}\gamma_{i,yr}^{HH}=\beta_{yr}+\gamma_{i}^{HH}$.
Population-level parameters -$\beta_{yr}$, $\gamma^{D}$ (death)
and $q$ - are updated exactly as described for the two-state model.

\subsection{North Atlantic humpback whale data}

The methods developed here are applied to a mark-resight data set
on a subpopulation of North Atlantic humpback whales sighted in the
Stellwagen Bank National Marine Sanctuary (SBNMS), in the Gulf of
Maine. Researchers from the Provincetown Centre for Coastal Studies
began documenting North Atlantic humpback whales in the Gulf of Maine
in 1975 and have to date individually identified over 1200. Humpback
whales (\emph{Megaptera novaeangliae}) are distributed worldwide,
with summer feeding ranges in mid to high-latitudes and winter breeding
in low-latitude areas \citep{Clapham1999}. They can be uniquely identified
by their natural markings: through the shape of their flukes and through
patterns from natural pigmentation \citep{Hammond1986}. \medskip{}

The Gulf of Maine is the southern most summer feeding ground for the
North Atlantic humpback whales. Individual humpback whales have been
intensively studied in this region since the late 1970s. The SBNMS
is one of several important feeding sites for North Atlantic humpback
whales which summer in the Gulf of Maine. Due to the consistent aggregation
of humpback whales and other marine life, the SBNMS was nominated
as a national sanctuary in 1992. The SBNMS encompasses only a small
part of the Gulf of Maine sub population's summer range, and although
some individuals are seen regularly there during the summer, none
are thought to remain permanently within its boundaries.

\section{Results\label{sec:Simulation-and-convergence} }

\subsection{Simulation results}

\subsubsection{Convergence with one long chain }

A simulated data set for 60 animals with capture history of length
500 was used as an initial exploratory test of the model. A MCMC chain
was run for 100000 iterations, with the first 10000 discarded to burn
in. Heterogeneity was included on each of the two hidden states and
on the probability of observation. Individual probabilities were drawn
from the following beta distributions: $\pi\sim Beta(4,2)$, $\gamma^{HH}\sim Beta(6,2)$,
$\gamma^{AA}\sim Beta(15,5)$.\medskip{}

\begin{figure}
\includegraphics[height=3in]{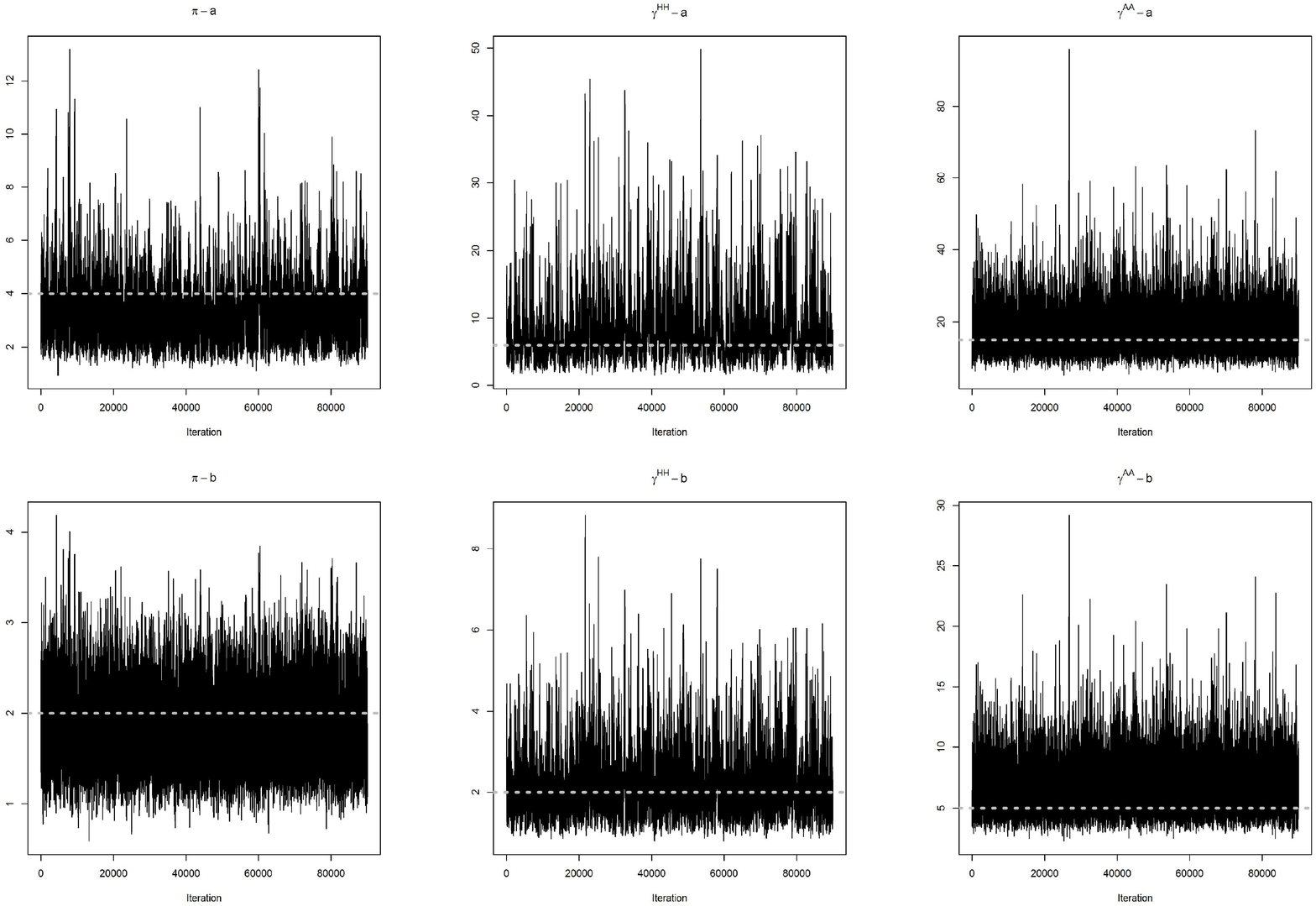}
\caption{Trace plots for population-level hyper-parameters. Dotted grey lines
indicate true values used in simulation of synthetic data. Plots indicate
convergence of chains.\label{fig:Figure1}}
\end{figure}

\begin{figure}
\includegraphics[height=3in]{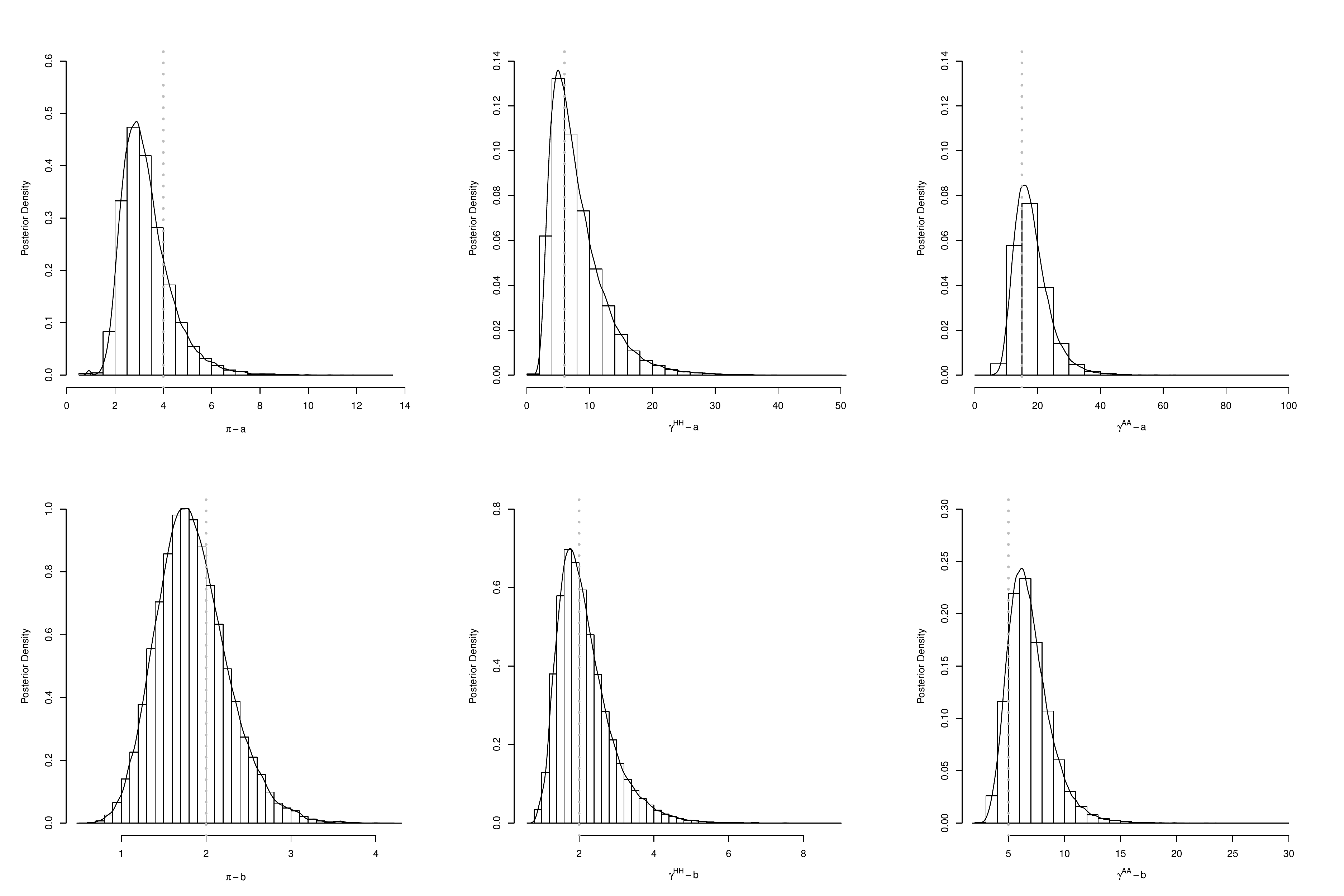}
\caption{Histogram and density plot of population-level hyper-parameters. Vertical
dashed lines indicate true value used to simulate data.\label{fig:Figure2}}
\end{figure}

\begin{figure}
\includegraphics[height=2in]{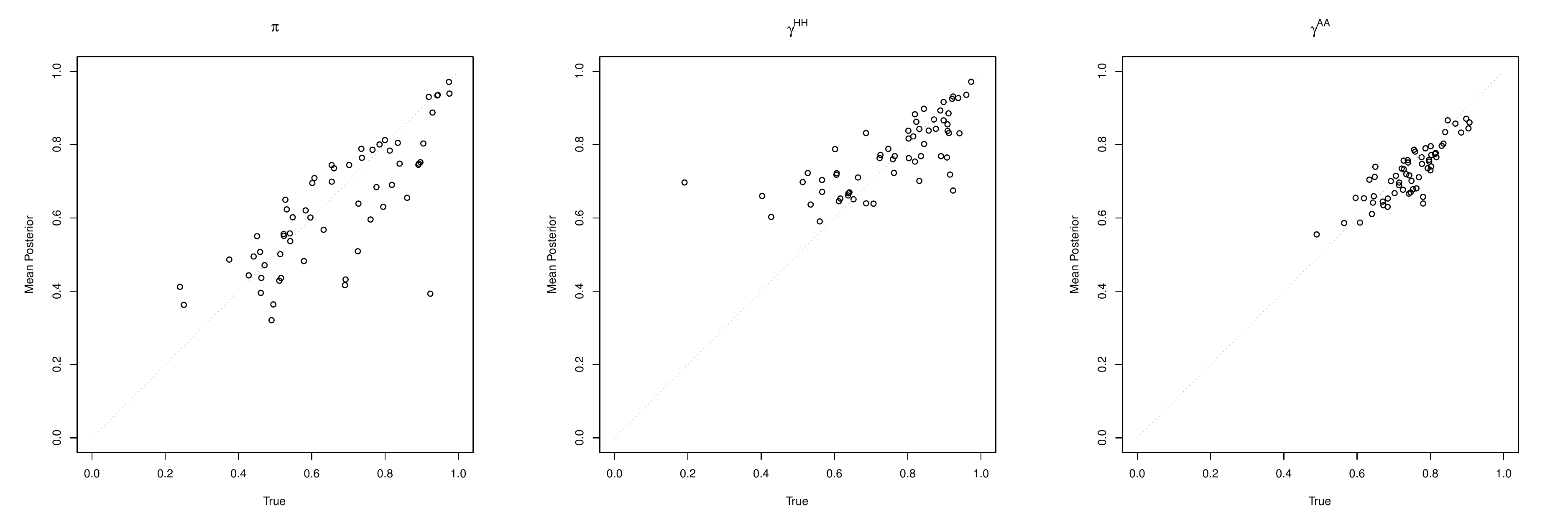}
\caption{Plot of true values from simulated data vs mean of individual posterior
values for probability of Observation, probability of remaining Here
and Away. \label{fig:Figure3} }
\end{figure}

Trace plots (Figure \ref{fig:Figure1}) indicate good
mixing for the six hyper-parameters and density plots (Figure \ref{fig:Figure2})
indicate convergence to true values. These parameters were updated
using the Independent Metropolis-Hastings algorithm; the mean acceptance
rate was 80\%. True values plotted against the mean posterior for
each individual for $\pi_{i}$, $\gamma_{i}^{HH}$ and $\gamma_{i}^{AA}$
(Figure \ref{fig:Figure3}) indicate strong correlation
between mean of individual posterior estimates and the true values
used in simulation.

\subsubsection{Asymptotic convergence of $a$ and $b$ }

The asymptotic convergence of the posterior distribution for each
of the hyper-parameters was tested using three simulated data sets
with fixed capture histories (length 250) but increasing number of
animals (50, 200, 800). For each parameter ($\pi$, $\gamma^{HH}$
and $\gamma^{AA}$), synthetic data was simulated from a $Beta(8,2)$
distribution (i.e. a mean of approximately 0.8). MCMC chains were
run for 15000 iterations, with the first 5000 discarded to burn in. 

\begin{table}
\resizebox{\textwidth}{!}{

\begin{tabular}{|c|c|c|c|c|c|c|}
\hline 
 & \multicolumn{2}{c|}{$\pi$} & \multicolumn{2}{c|}{$\gamma^{HH}$} & \multicolumn{2}{c|}{$\gamma^{AA}$}\tabularnewline
\hline 
\hline 
 & $a$ & $b$ & $a$ & $b$ & $a$ & $b$\tabularnewline
\hline 
TRUTH & 8 & 2 & 8 & 2 & 8 & 2\tabularnewline
\hline 
\hline 
50 animals & 13.17 ( \textbf{6.41 }) & 2.71 ( \textbf{1.00} ) & 7.34 ( \textbf{2.14} ) & 1.93 ( \textbf{0.45} ) & 10.40 ( \textbf{2.60} ) & 2.33 (\textbf{ 0.53} )\tabularnewline
\hline 
200 animals & 9.47 ( \textbf{1.64 }) & 2.29 ( \textbf{0.34} ) & 7.07 ( \textbf{1.17} ) & 1.77 ( \textbf{0.24} ) & 6.60 ( \textbf{0.94} ) & 1.65 ( \textbf{0.19} )\tabularnewline
\hline 
800 animals & 7.98 ( \textbf{0.88} ) & 1.97 ( \textbf{0.17 }) & 8.62 ( \textbf{0.93} ) & 2.17 ( \textbf{0.17} ) & 8.77 ( \textbf{0.68 }) & 2.08 ( \textbf{0.13} )\tabularnewline
\hline 
\end{tabular}}

\caption{Means and standard deviations (in brackets) for hyper-parameters from
simulated data. \label{tab:Means-and-standard}}
\end{table}
\medskip{}

\begin{figure}
\includegraphics[height=3in]{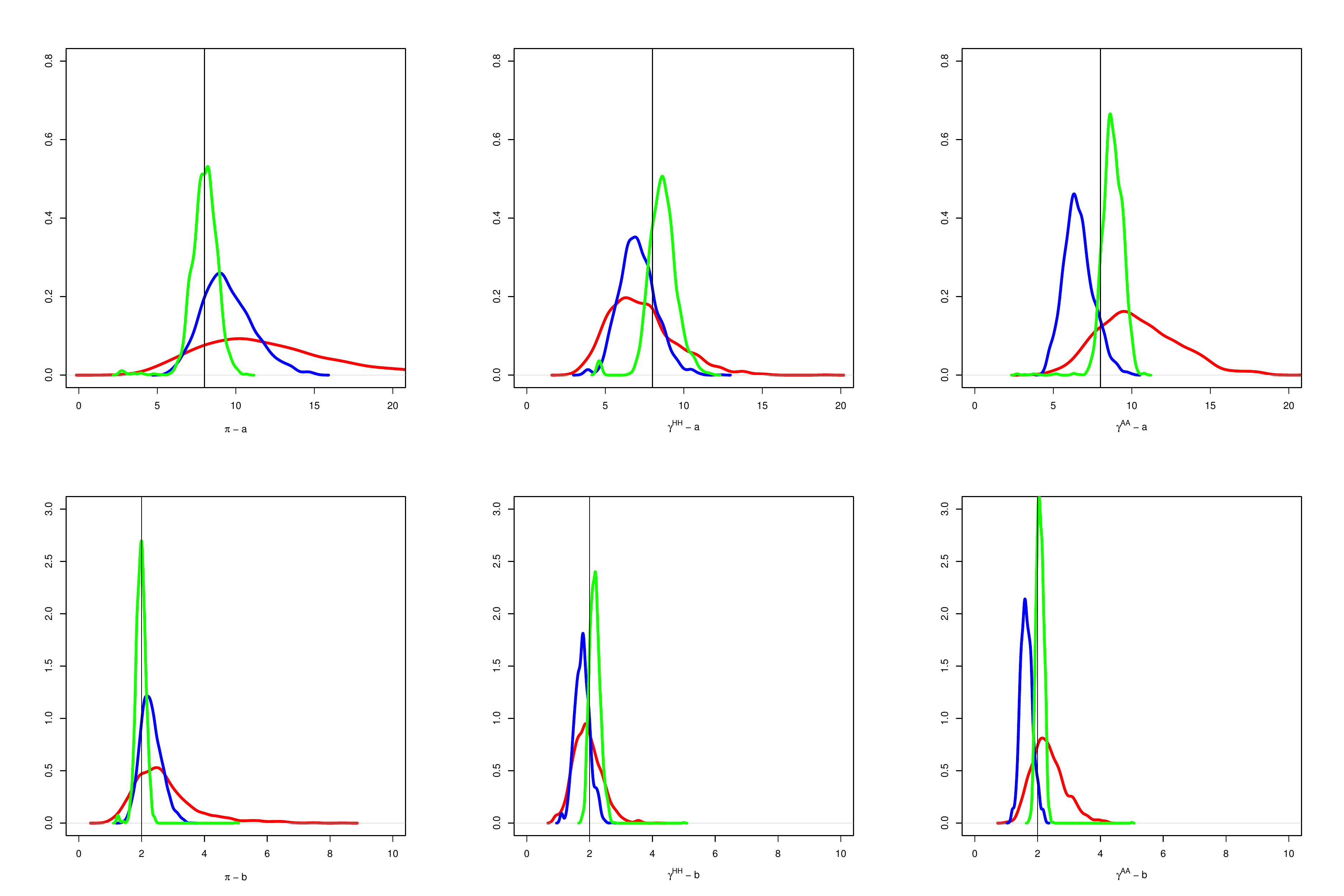}
\caption{Posterior density plots for each hyper-parameter from simulated data.
The red density line is for 50 animals; blue for 200 animals; and
green for 800 animals. The vertical black line indicates the true
values used to simulate the data. \label{fig:Figure4} }
\end{figure}

Figure \ref{fig:Figure4} shows density plots of posterior
estimates for the six hyper-parameters: decreasing variance is evident
with increasing number of individuals. Table \ref{tab:Means-and-standard}
indicates posterior standard deviations decreased at the expected
rate of approximately $\frac{1}{\sqrt{n}}$ (here $n$ = number of
animals). 

\subsubsection{Convergence of individual mean posterior estimates }

Asymptotic convergence for the mean of individual posterior estimates
to true individual values (used in simulation) was tested by increasing
length of capture history (250, 1000, 4000) for three data sets each
with 16 individuals. MCMC chains were run for 15000 iterations, with
the first 5000 discarded to burn in. Data were simulated using the
following distributions: $\pi\sim Beta(30,3)$, $\gamma^{HH}\sim Beta(30,5)$
and $\gamma^{AA}\sim Beta(30,2)$ (i.e. approximate mean probabilities
of 0.91, 0.86, 0.94). \medskip{}

\begin{figure}
\includegraphics[height=2in]{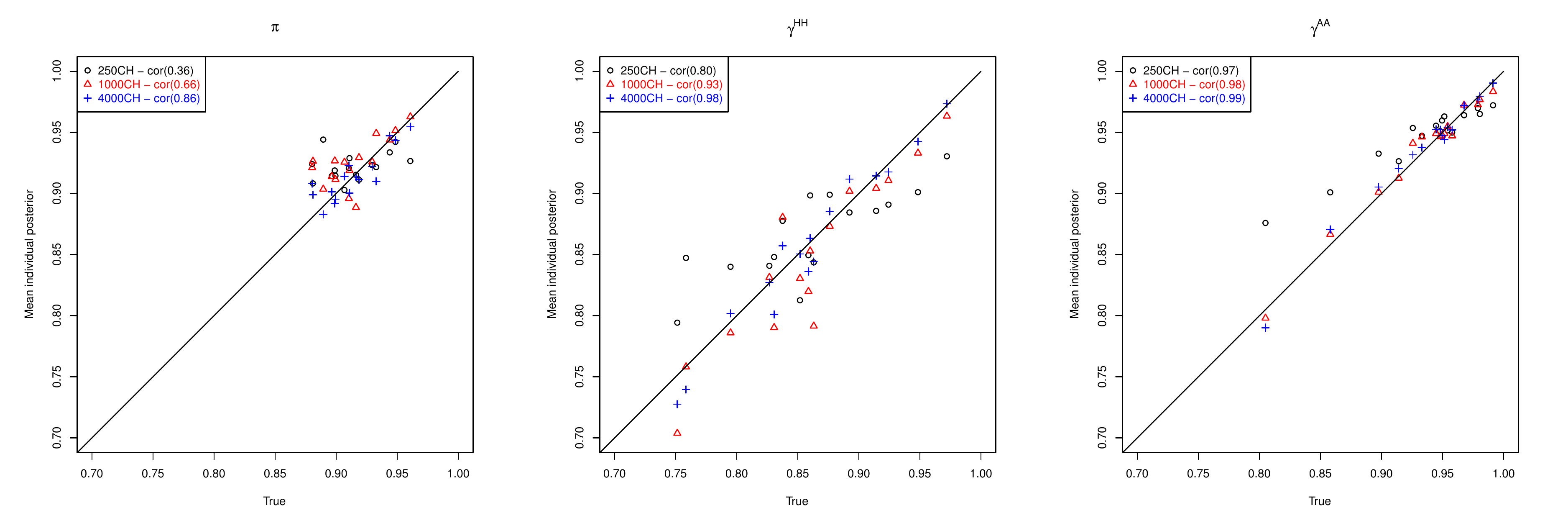}
\caption{Mean individual posterior values vs true values for $\pi$, $\gamma^{HH}$
and $\gamma^{AA}$ from simulated data. Correlations indicate increasing
convergence to true values with increasing length of capture history.
\label{fig:Figure5}}
\end{figure}

Figure \ref{fig:Figure5} indicates convergence of the mean
of individual posterior estimates to true values for each of $\pi$,
$\gamma^{HH}$ and $\gamma^{AA}$. Standard deviations decreased with
increasing length of capture history, and correlation improved with
longer capture histories. 

\medskip{}

\subsection{Results for North Atlantic Humpback whales}

\begin{figure}
\includegraphics[height=3in]{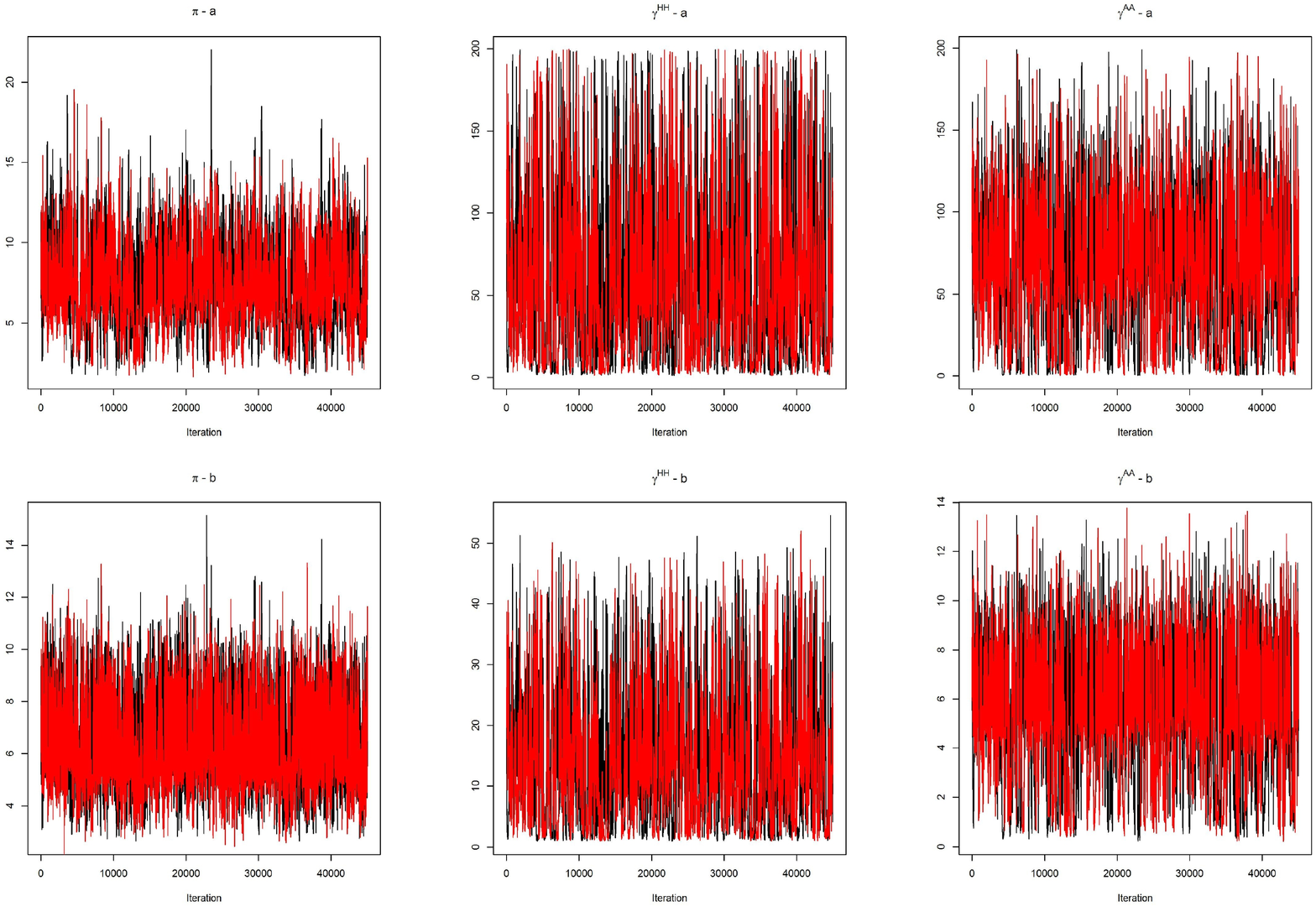}
\caption{Trace plots of hyper-parameters for $\pi$, $\gamma^{HH}$ and $\gamma^{AA}$
for 90000 posterior estimates for real data. The black trace indicates
the first 45000 estimates and the red the final 45000.\label{fig:Figure6} }
\end{figure}

\begin{figure}
\includegraphics[height=3in]{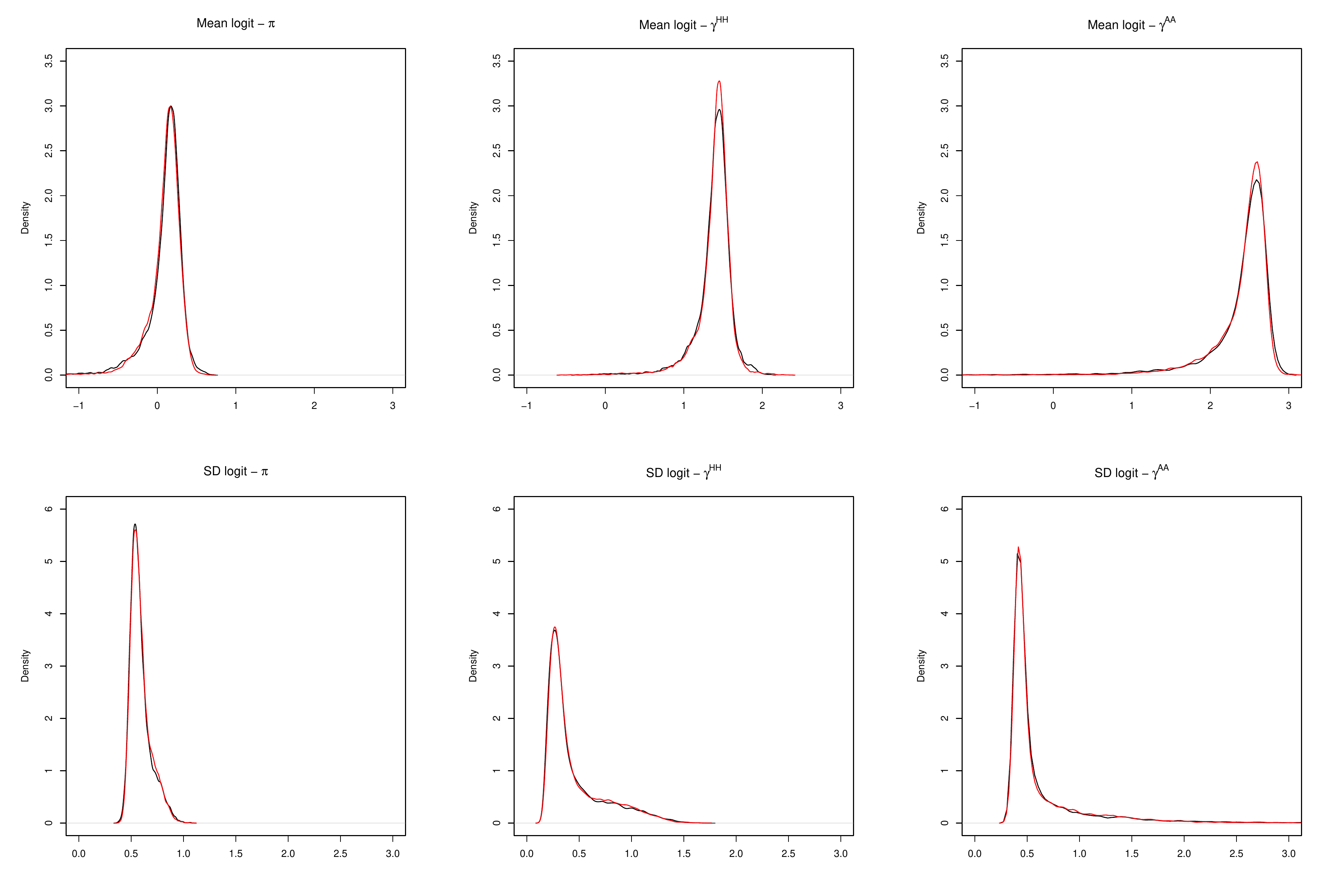}
\caption{Density plots for the mean and standard deviation of logit$\pi$,
$\gamma^{HH}$ and $\gamma^{AA}$ for 90000 posterior estimates for
real data. The black line indicates the first 45000 estimates and
the red the final 45000. \label{fig:Figure7}}
\end{figure}

One chain was run for 100000 iterations with the first 10000 discarded
to burn-in. Figure \ref{fig:Figure6} indicates the trace
plots for the first and last 45000 posterior estimates and Figure
\ref{fig:Figure7} shows density plots of the mean and
standard deviation of logit $\pi$, $\gamma^{HH}$ and $\gamma^{AA}$
(calculated as $mean=\psi(a)-\psi(b)$ and $sd=\sqrt{\psi(a)'+\psi(b)'}$
where $\psi$ indicates the moment). Both the trace plots and a the
mean and standard deviation on the logit scale indicate no visible
differences between the first and second half of the chains suggesting
that full convergence was reached with 100000 iterations. \medskip{}

\begin{figure}
\includegraphics[height=2in]{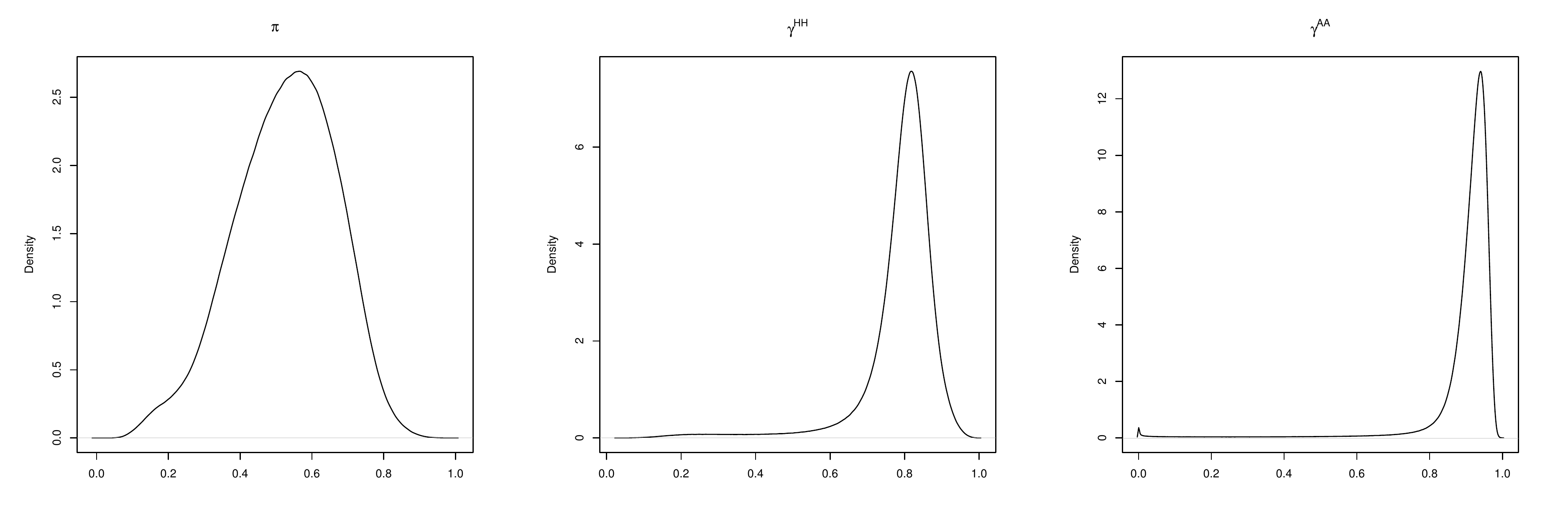}
\caption{Density plots from 90000 posterior samples from three independent
chains for $\pi$, $\gamma^{HH}$ and $\gamma^{AA}$ for real data.
\label{fig:Figure8}}

\end{figure}

Posterior estimates were less varied for $\pi$ compared to $\gamma^{HH}$
and $\gamma^{AA}$. Figure \ref{fig:Figure8} indicates the
density for $\pi$, $\gamma^{HH}$ and $\gamma^{AA}$ for all individuals
for 90000 iterations. Population-level variability is apparent in
$\pi$, but somewhat less so in the two state probabilities. However,
Figure \ref{fig:Figure7} indicates that the standard deviation
logits are not much smaller for both $\gamma^{HH}$ and $\gamma^{AA}$
compared to $\pi$. Since the transition parameters ($\gamma^{HH}$
and $\gamma^{AA}$) are close to 1, there is less opportunity to see
heterogeneity than for $\pi$. 

\subsubsection{Fully Bayesian MCMC approach vs Empirical Bayes using ADMB}

The same subset of whale (176 individuals seen more than 30 times
between 1979 and 2005) was used for the Empirical Bayes model in ADMB.
Beta distributed random effects were implemented in ADMB. The model
was otherwise the same as described in \citet{Ford2012}. 

\begin{figure}
\includegraphics[height=2in]{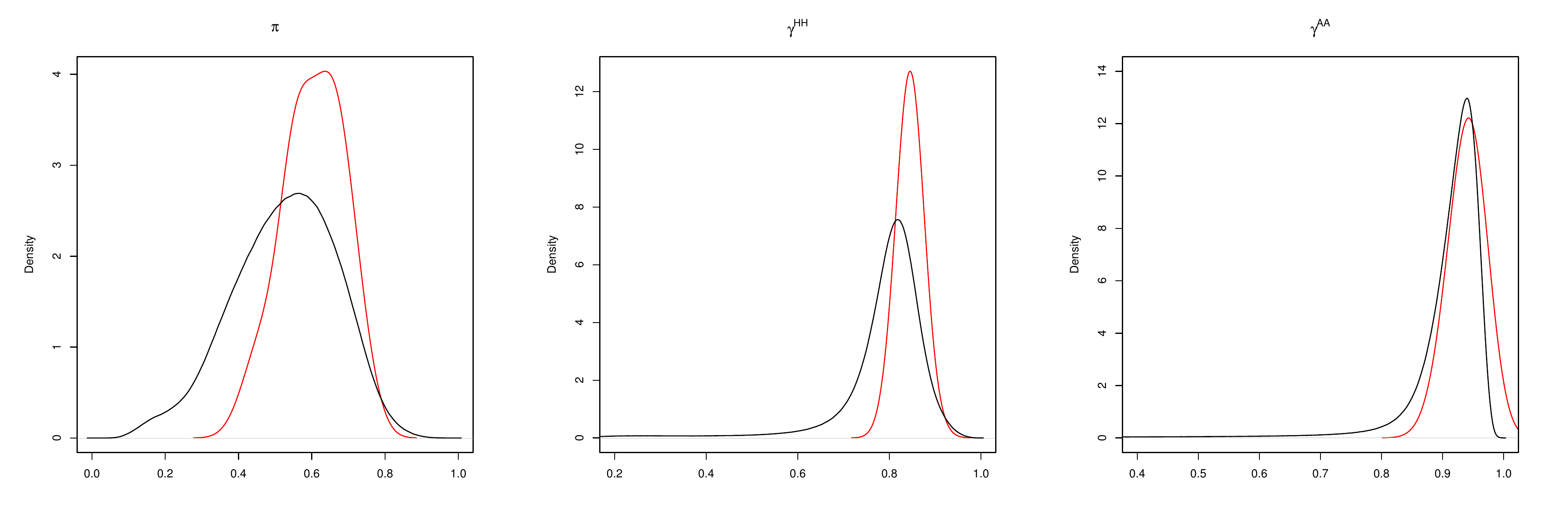}
\caption{Comparison of posterior samples for $\pi$, $\gamma^{HH}$ and $\gamma^{AA}$
for Beta-Binomial MCMC sampler and individual posterior estimates
from ADMB. \label{fig:Figure9}}

\end{figure}

Posterior estimates from the beta-binomial MCMC model were compared
with individual posterior results from ADMB. Results appear to be
similar, but more variation and slightly lower mean is evident in
the Beta-Binomial MCMC results (Figure \ref{fig:Figure9}).
This difference in variation is likely due to the Beta-Binomial MCMC
results incorporating the uncertainty on the hyper-parameters; in
comparison ADMB results are conditioned on point estimates of the
hyper-parameters. 

\section{Discussion\label{sec:Discussion}}

We have developed a hierarchical Bayesian hidden Markov model where
individual parameters are updated using Gibbs sampling, the population-level
hyper-parameters and fixed-effects using an Independent Metropolis-Hastings
sampler. Results from simulation tests indicate asymptotic convergence
for both individual and population-level parameters: the posterior
standard deviations for the population-level parameters decreased
with increasing number of individuals, and correlation for mean posterior
individual estimates to true values improved with increasing length
of capture history.\medskip{}

The Beta-binomial model we presented in this paper is an efficient
MCMC routine for incorporating individual heterogeneity into mark-recapture
models. The Independent sampler in particular provides an efficient
method to update parameter values when a Gibbs sampler cannot be implemented.
The Independent sampler is almost efficient as the Gibbs sampler and
avoids the difficulties which can arise with standard Metropolis-Hastings
routines, namely the need to tune random walk sizes. \medskip{}

Output from ADMB was used to facilitate sampling from a multivariate
t-distribution in order to update the fixed parameters in the model.
This resulted in an automatic and computationally efficient method
to update these complex model parameters, avoiding the need to tweak,
for example, random walk step sizes. This method can also be used
to incorporate covariates into the model, providing another extension
to the already flexible model. We note here that we used Laplace Approximation
results from ADMB in order to inform the fully Bayesian approach using
the Beta-binomial MCMC sampler. This provided an efficient and effective
method to update the MCMC sampler. \medskip{}

In Empirical Bayes methods (using Laplace Approximation), the hyper-parameter
is estimated from the data and set to it's maximum likelihood estimate.
In comparison, in a fully Bayes approach the hyper-parameter is endowed
with a prior distribution. Thus the MCMC approach automatically incorporates
uncertainty into the hyper-parameters. We presented results from a
beta-binomial model implemented purely in ADMB. The ADMB software
does include an MCMC option, however our experience was that it ran
extremely slowly before eventually crashing due to memory constraints.
Thus we were unable to obtain full MCMC results from ADMB (using version 11). If this situation is rectified in future releases
of the software a full Bayesian version of the approach may avoid
the need to resort to MCMC. While our experience was that MCMC was
quicker and worked, it did require significant use of ADMB estimation
in order to inform the MCMC. This demonstrates that in real situations,
hybrid approaches can be particularly useful in getting complex multi-state
models to perform reliably. 

\medskip{}

The approach we present in this paper, presents a tractable Bayesian
method to estimate multi-state mark-recapture models which incorporate
individual heterogeneity on process and observation model parameters.
As MCMC samplers go it is computationally efficient and therefore
allows for thorough model checking via simulation. Our application
to data from humpback whales indicated heterogeneity in both site
fidelity and sighting probability. This indicates heterogeneity in
propensity to use the marine sanctuary both on a population and individual-level.
The role of individuals in mediating population processes has been
a key question in ecology but it is often one that is side-stepped
in management and conservation applications. In this paper we have
demonstrated an analytical tool which we hope will begin to fill this
gap. 

\medskip{}

\section*{Acknowledgements}
We thank Chris Wilcox and Jooke Robbins for much intellectual input and discussion, and Jooke Robbins and the Provincetown Center for Coastal Studies for data. 

\bibliographystyle{humanbio}
\bibliography{beta}
\end{document}